# A Kullback–Leibler Divergence-based Distributionally Robust Optimization Model for Heat Pump Day-ahead Operational Schedule in Distribution Networks

Zihao Li, Wenchuan Wu, *Senior Member, IEEE,* Boming Zhang, *Fellow, IEEE*

***Abstract*—For its high coefficient of performance and zero local emissions, the heat pump (HP) has recently become popular in North Europe and China. However, the integration of HPs may aggravate the daily peak-valley gap in distribution networks significantly. In this paper, we describe a distributionally robust optimization (DRO)-based heat pump day-ahead operational schedule model (HP-DOSM) to shave the peak and reduce residents' costs under time-of-use. The ambiguity set of the DRO is constructed using Kullback–Leibler divergence with a nominal distribution. This model can well capture the uncertainties of weather, photovoltaic, and load prediction errors. Moreover, this DRO based HP-DOSM can be transformed into a tractable deterministic model. Compared with robust optimization (RO) models, our model is less conservative since more statistical information of the uncertainties is utilized. Numerical tests were conducted to demonstrate its performance, compared with the RO model via Monte Carlo simulations.**

## Nomenclature

| | |
|---|---|
| $N$ | Total number of houses equipped with heat pump and water tank |
| $H$ | Time horizon in 1 day |
| $T_{k,t}$ | Indoor air temperature of the $k^{th}$ house at time $t$ |
| $T_{w,k,t}$ | Tank water temperature of the $k^{th}$ house at time $t$ |
| $T_{out,t}$ | Outdoor temperature at time $t$ |
| $T_k$ | Vector $[T_{k,0} \ T_{k,1} \ \cdots \ T_{k,H}]^T$ |
| $T_{w,k}$ | Vector $[T_{w,k,0} \ T_{w,k,1} \ \cdots \ T_{w,k,H}]^T$ |
| $T_{out}$ | Vector $[T_{out,0} \ T_{out,1} \ \cdots \ T_{out,H}]^T$ |
| $\overline{T}_k$ | Upper temperature boundary of the $k^{th}$ house |
| $\underline{T}_k$ | Lower temperature boundary of the $k^{th}$ house |
| $R_k$ | Thermal resistance of the $k^{th}$ house |
| $C_k$ | Thermal capacitor of the $k^{th}$ house |
| $R_{w,k}$ | Thermal resistance of the $k^{th}$ water tank |
| $C_{w,k}$ | Thermal capacitor of the $k^{th}$ water tank |
| $x_{k,t}$ | ON/OFF (1/0) state of the $k^{th}$ heat pump |
| $x_k$ | Vector $[x_{k,0} \ x_{k,1} \ \cdots \ x_{k,H}]^T$ |
| $P_{HP,k}$ | Rated power of the $k^{th}$ heat pump |
| $\gamma_{COP,k}$ | Coefficient of performance of the $k^{th}$ HP |
| $\gamma_{w2h,k}$ | Water tank to house heating efficiency of the $k^{th}$ house |
| $Z$ | Zone of local transformer and feeder lines |
| $P_{Load,k,t}$ | Load of the $k^{th}$ house except heat pump at time $t$ |
| $P_{PV,k,t}$ | Photovoltaic power of the $k^{th}$ house at time $t$ |
| $P^Z_{trans,t}$ | Transformer power of zone $Z$ at time $t$ |
| $\xi_{P,t}$ | Forecast errors of power at time $t$ |
| $\xi_{T,t}$ | Forecast errors of outdoor temperature at time $t$ |
| $P_t$ | Distribution of power forecast error at time $t$ |
| $\mathrm{P}_t$ | Ambiguity set of $P_t$ |
| $P_{0t}$ | Center/nominal distribution of $\mathrm{P}_t$ |
| $Q_t$ | Distribution of outdoor temperature forecast error at time $t$ |
| $\mathrm{Q}_t$ | Ambiguity set of $Q_t$ |
| $Q_{0t}$ | Center/nominal distribution of $\mathrm{Q}_t$ |
| $\eta_t$ | Farthest KL divergence in the ambiguity set of power forecast error at time $t$ |
| $\chi_t$ | Farthest KL divergence in the ambiguity set of outdoor temperature forecast error at time $t$ |

## I. Introduction

Heat pumps (HPs) have seen a revival in recent years, in which their potential role with renewable energy sources is being investigated. HP technologies have experienced great improvements in operating stability and have subzero coefficient of performance [1], providing confidence for consumers in their reliability. Unsurprisingly, more than 750,000 units were sold in the EU-20, reducing greenhouse gas emissions by 6.8 Mt in 2010 [2]. However, HPs have placed a considerable load on power systems, although recognized as a promising resource for demand-side management in using renewable energy [3]. In recent years, a policy named 'Coal to Electricity' has been promulgated in Beijing, to replace coal stoves in the nearby countryside with HPs for heating, which may be beneficial in reducing the gray smog sky over Beijing [4]. By the end of November 2016,

Manuscript received XXX, 2017. This work was supported in part by the National Science Foundation of China (51477083), in part by Key Science and Technology Project of State Grid Corporation of China (Grant. 5202011600U4). The authors are with the State Key Laboratory of Power Systems, Department of Electrical Engineering, Tsinghua University, Beijing 100084, China (e-mail: wuwench@tsinghua.edu.cn).

227,000 houses in 663 counties in Beijing had had electrical heaters installed [5]. However, the local distribution networks take the risk of having insufficient capacity in transformers and branches, where more investment in infrastructure is needed. An optimal day-ahead operational schedule strategy for HPs may provide the benefit of peak-shaving [6] and make better use of renewable energy [7].

For a local distribution network, integrated with renewable energy generators and HPs, its day-ahead operational schedule suffers uncertainties, from, for example, weather predictions, load predictions, and renewable energy predictions. Conventionally, stochastic optimization (SO) and robust optimization (RO) techniques have been used to capture uncertainties. Recently, distributionally robust optimization (DRO) has been considered a more practical paradigm for decision making under uncertainty, where the uncertain variable is governed by a probability distribution type that is itself subject to uncertainty [8]. DRO provides more robustness with ambiguous distribution sets and is less conservative than classical RO; moreover, it does not require explicit distribution of uncertain variables, like SO, and demonstrates application potential in many areas [9-12].

The ambiguity set of a DRO problem is a family of distributions, characterized through certain known properties of the real distribution [13] and must be rich enough to include the real distribution while small enough to exclude pathological distributions [14]. How the ambiguity set is constructed is key to DRO, and there are two main ways: moments-based and distance-based methods. The moments-based method usually defines an ambiguity set by restricting the mean and covariance matrix of the distributions to some given values; thus, the DRO model can be further transformed into a deterministic Semidefinite Programming(SDP) or Second Order Programming(SOCP) problem under several linear assumptions and approximations [12, 15, 16]; however, odd distributions and mixed integer problems may introduce additional complexity. Distance-based methods define the ambiguity set as a ball in the measured space of probability distributions, using a probability distance function, such as the Prokhorov metric [17], Kullback–Leibler divergence (KL divergence) [15, 18], or the Wasserstein metric [14]. Distance-based ambiguity sets limits on the shape of distributions that explicitly use data distribution information and may decrease the conservatism and the robustness that could be controlled with explicit meanings.

Due to the ON/OFF nature of HPs, the day-ahead HP operational schedule is a mixed-integer linear program (MILP) with uncertain variables. For moments-based DRO models, methods such as Benders decomposition [19], linear approximation [11], and conic programming methods [20], are used to solve them. However, if these models incorporate mixed-integer variables, they may become intractable.

In this paper, a KL distance-based MILP DRO model is described for a HP day-ahead operational schedule. It can be transformed to a tractable deterministic MILP model. Moreover, the distance-based DRO can better use historical data [14, 15, 21]. In [15], it was suggested that adopting distance-based ambiguity sets is less conservative.

This paper is based on the works [15, 18], and the major contributions include the following:

(1) A heat pump day-ahead operational schedule model (HP-DOSM) is proposed to shave the peak and reduce residents' cost, which is formulated as a MILP DRO. A KL divergence-based ambiguity set is used to capture uncertainties in the prediction errors in outdoor temperature, residential load, and photovoltaic power.

(2) The proposed MILP DRO model is transformed to a tractable deterministic MILP problem under two nominal distributions: a Gaussian distribution and a kernel density estimator (KDE).

(3) Numerical tests show that the proposed approach has potential to provide effective peak shaving and outperforms the conventional RO model.

The remainder of the paper is organized as follows. Section II presents the deterministic day-ahead HP operational schedule model. Section III discusses the formulation of MILP DRO model and use of historical data. Numerical tests are described in Section IV. The paper ends with conclusions and future work in Section V.

## II. DETERMINISTIC HEAT PUMP DAY-AHEAD OPERATIONAL SCHEDULE model

Figure 1 shows a local distribution network integrated with photovoltaic panels and consumers with air-source HPs. For comfort considerations, the HP first heats the water in a tank to supply heat for the house. The tank is installed indoors and links to a finned tube with constant water flux.

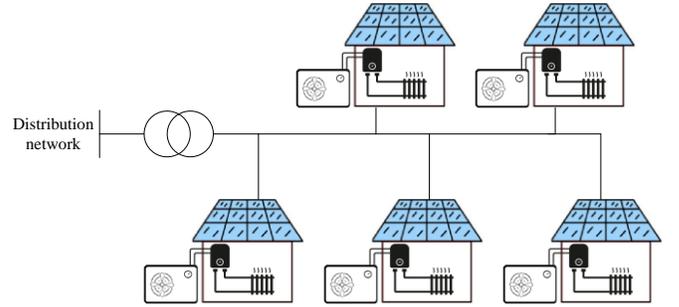

Figure 1. Local distribution network

We apply a one-order equivalent thermal RC model to simulate the house and the water tank. The RC model is formulated as follows,

$$x_k \gamma_{COP,k} P_{HP,k} + \frac{T_k - T_{w,k}}{R_{w,k}} = C_{w,k} \dot{T}_{w,k} \tag{1}$$

$$\frac{T_{out} - T_k}{R_k} + \gamma_{w2h,k} \frac{T_{w,k} - T_k}{R_{w,k}} = C_k \dot{T}_k \tag{2}$$

where $R_k, C_k$ and $R_{w,k}, C_{w,k}$ are the 1st-order thermal parameters of the house and water tank, respectively, of the $k^{th}$ house, and they can be identified from historical measurements for real application. $T_k$, $T_{w,k}$, and $T_{out}$ are the indoor air, water tank, and outdoor air temperatures, respectively. $P_{HP,k}$, $\gamma_{COP,k}$, $\gamma_{w2h,k}$, and $x_k$ are the rated power, constant HP to water coefficient of performance (COP), constant water-to-house efficiency, and the ON-OFF state of the $k^{th}$ houses's HP. The difference schemes of (1) and (2) are

$$T_{k,t+1} = \frac{R_k(T_{w,k,t} - T_{k,t})}{R_{w,k}} + T_{out,t}$$

$$+ \{T_{k,t} - [T_{out,t} + \frac{\gamma_{w2h,k} R_k(T_{w,k,t} - T_{k,t})}{R_{w,k}}]\} e^{-\frac{\Delta t}{R_k C_k}} \quad (3)$$

$$T_{w,k,t+1} = R_{w,k} \gamma_{COP,k} P_{HP,k} x_{k,t} + T_{k,t}$$

$$+ (T_{w,k,t} - (R_{w,k} \gamma_{COP,k} P_{HP,k} x_{k,t} + T_{k,t})) e^{-\frac{\Delta t}{R_{w,k} C_{w,k}}} \quad (4)$$

where $\Delta t$ is the time interval.

The HP operates in two alternative modes: OFF and ON. Thus, the power consumed by a HP is a function of its operational state:

$$P_{HP,k,t} = x_{k,t} P_{HP,k}, \text{ where } x_{k,t} = \begin{cases} 1, \text{ for } state_{k,t} = ON \\ 0, \text{ for } state_{k,t} = OFF \end{cases} \quad (5)$$

The objective function of the HP day-ahead operational schedule we propose includes shaving the peak load at the local transformer and minimizing the electricity cost:

$$\text{minimize } \psi \sum_{Z \in Trans} P_{max}^Z + \sum_{k \in Z} \sum_t \rho_t x_{k,t} P_{HP,k} \quad (6)$$

where $P_{max}^Z$ represents one-day's maximum power load on the transformer $Z$, $\psi$ is a penalty weight for peak shaving, called the peak cost factor [22], and $\rho_t$ represents the time of use (TOU) electricity price.

A DC power flow model is used for the distribution network and the power balance constraint for the transformer of zone $Z$ is

$$P_{Trans,t}^Z = \sum_{k \in Z} x_{k,t} P_{HP,k} + \sum_{k \in Z} P_{Load,k,t} - \sum_{k \in Z} P_{PV,k,t} \quad (7)$$

where $P_{Load,k,t}$ and $P_{PV,k,t}$ are the residential loads (except the HPs) and the PV output.

The capacity constraints of the transformer are

$$P_{max}^Z \geq P_{Trans,t}^Z, t = 1, 2, ..., H \quad (8)$$

$$P_{max}^Z \leq P_{trans\_capacity}^Z \quad (9)$$

where $P_{trans\_capacity}^Z$ is the rated capacity of the local transformer $Z$ and $H$ is the total schedule time periods of 1 day.

Thermal constraints include (3), (4), indoor temperature comfort constraints:

$$\underline{T_k} \leq T_{k,t} \leq \overline{T_k} \quad (10)$$

The initial temperature settings are

$$T_{k,t=0} = T_{k,t=0,set} \quad (11)$$

$$T_{w,k,t=0} = T_{w,k,t=0,set} \quad (12)$$

The daily water temperature hold is

$$T_{w,k,t=H} \geq T_{w,k,t=0} \quad (13)$$

and the HP ON-OFF switching time-delay limitations are

$$-2 < 2x_{k,t} - x_{k,t-1} - x_{k,t+1} < 2 \quad (14)$$

where index $t$ represents the schedule time, $T_{k,t=0,set}$ and $T_{w,k,t=0,set}$ are the initial temperatures of the indoor air and the tank water, and $\overline{T_k}, \underline{T_k}$ are the indoor air temperature comfort range bounds. The switching time-delay constraints indicate that the HP needs to keep the same mode for at least two time intervals.

Then, the deterministic HP-DOSM can be formulated as:

*Model 1: Deterministic heat pump day-ahead operational schedule model*

$$\min \ (6)$$
$$s.t. \ (3)(4)(7)-(14)$$
$$x_{k,t} \in \{0,1\}, \forall t = 1, \cdots, H, \forall k = 1, 2, \cdots, N, \forall Z$$

With the given prediction data, this MILP model can solved efficiently. However, the schedule decision based on bias prediction may lead to infeasibility in real operation. Accordingly, we introduce a KL divergence-based DRO in the next section, considering the uncertainties of forecast errors in outdoor temperature, load prediction, and PV prediction.

## III. KL DIVERGENCE-BASED DRO MODEL FOR HP DAY-AHEAD OPERATIONAL SCHEDULE

In this section, we first present the KL divergence-based DRO (*KL-DRO*) formulation to minimize total cost given the expectation constraints. Then, this *KL-DRO* model is transformed to a deterministic MILP. Finally, we discuss how to use historical data to build an ambiguity set for the *KL-DRO* model as well as the risk meaning of the radius distance of the ambiguity set.

There are several statistical distance functions, such as the Prokhorov metric, the KL divergence, and the Wasserstein metric, which show common properties and can co-transfer in some situations [17]. Here, the KL divergence is adopted and is defined as

$$D_{KL}(p(\theta) \| q(\theta)) = \int_\Omega p(\theta) \ln \frac{p(\theta)}{q(\theta)} d\theta \quad (15)$$

where $p, q$ are distribution functions in measure space $\Omega$.

### A. KL-DRO model

To capture the uncertainties in prediction errors, we reform the deterministic constraints, (8), (10), and (13), as expected constraints under the worst distribution in the ambiguity sets: i.e., constraints (7) and (8) are reformed as

$$\max_{P \in \mathbb{P}_t} \sum_{k \in Z} x_{k,t} P_{HP,k} + P_{sum,t}^Z - P_{max}^Z \leq 0, t = 1, 2, \cdots, H \quad (16)$$

Here, $P_{sum,t}^Z = \sum_{k \in Z} P_{Load,k,t} - \sum_{k \in Z} P_{PV,k,t}$ is the summation of load demand and PV output in zone Z. $P_{sum,t}^Z$ is further expressed as

$$P_{sum,t}^Z = \hat{P}_{sum,t}^Z + \xi_{P,t} \quad (17)$$

where $\hat{P}_{sum,t}^Z$ is the prediction value and $\xi_{P,t}$ represents the prediction error.

The distribution function $P_t$ of $\xi_{P,t}$ belongs to the ambiguity set $\mathbb{P}_t$, defined as

$$\mathbb{P}_t = \{P_t \in \mathbb{P}_t \mid D_{KL}(P_t \| P_{0t}) \leq \eta_t\} \quad (18)$$

where $\eta_t$ is the farthest KL divergence with the nominal distribution $P_{0t}$, also referred to as the radius.

In a similar way, we can also reformulate constraints (10) and (13) as

$$\max_{Q\in Q} E_Q[J_k x_k + K_k T_{out} + l_k - \overline{T}_k] \leq 0 \quad (19)$$

$$\max_{Q\in Q} E_Q[-J_k x_k - K_k T_{out} - l_k + \underline{T}_k] \leq 0 \quad (20)$$

$$\max_{Q\in Q} E_Q\left[Z_0(M_{w,k} x_k + N_{w,k} T_{out} + p_{w,k}) + T_{w,k,t=0,set}\right] \leq 0 \quad (21)$$

where $x_k$, $T_k$, $T_{w,k}$, $T_{out}$ are the vectors explained in the nomenclature, $Z_0 = \begin{bmatrix} 0 & 0 & \cdots & -1 \end{bmatrix}$, and the equalities (3) and (4) are merged in these formulae, $T_{out,t} = T_{out,t} + \xi_{T,t}$, for dividing the prediction value and the uncertain part. The distribution function $Q_t$ of $\xi_{T,t}$ belongs to the ambiguity set $Q_t$, which is defined as $Q_t = \{Q_t \in Q_t \mid D_{KL}(Q_t \| Q_{0t}) \leq \chi_t\}$ and $\chi_t$ is the farthest KL divergence (radius) with the nominal distribution $Q_{0t}$. Details of this reformulation are provided in Appendix A.

Thus, the *KL-DRO* model for the HP day-ahead operational schedule is described as follows.

*Model 2: KL-DRO model*
$$\min \quad (6)$$
$$s.t. \quad (9),(11)(12)(14),(16)\text{-}(21)$$
$$x_{k,t} \in \{0,1\}, \forall t = 1,\cdots,H, \forall k = 1,2,\cdots,N, \forall Z$$

Selection of the nominal distribution and the radius will be further discussed in subsections *B* and *D*.

*B. Reformulation*

The nominal distribution $P_0$ of the ambiguity set contains all of the information driven from historical data. The historical data are the deviation records of predicted versus real values.

Referring to [18], a DRO expectation constrained program under KL divergence ambiguity such as

$$\min_{x\in X} h(x)$$
$$s.t. \max_{P\in P} E_P[H(x,\xi)] \leq 0 \quad (22)$$
$$P = \{P \in P \mid D_{KL}(P \| P_0) \leq \eta\}$$

can be transformed into the following form on the basis of strong duality theory,
$$\min_{x\in X} h(x)$$
$$s.t. \min_{\alpha \geq 0}\{\alpha \ln E_{P_0}[e^{H(x,\xi)/\alpha}] + \alpha\eta\} \leq 0 \quad (23)$$

if for every $x \in X$, $S = \{s \in \Re \mid s > 0, E_{P_0}[e^{sH(x,\xi)}] < +\infty\}$ is not empty.

Based on the theorem, the expectation constraints, (16)-(21), in *Model 2* can be transformed into
$$\min_{\alpha\geq 0}\{\alpha \ln E_{P_0}[e^{[\sum_{k\in Z} x_{k,t} P_{HP,k} + P^Z_{sum,t} - P^Z_{max} + \xi_{P,t}]/\alpha}] + \alpha\eta\} \leq 0 \quad (24)$$

$$\min_{\alpha\geq 0}\{\alpha \ln E_{Q_0}[e^{[J_k x_k + K_k T_{out} + l_k - \overline{T}_k + K_k \xi_T]/\alpha}] + \alpha\chi\} \leq 0 \quad (25)$$

$$\min_{\alpha\geq 0}\{\alpha \ln E_{Q_0}[e^{[-J_k x_k - K_k T_{out} - l_k + \underline{T}_k - K_k \xi_T]/\alpha}] + \alpha\chi\} \leq 0 \quad (26)$$

$$\min_{\alpha\geq 0}\{\alpha \ln E_{Q_0}[e^{[Z_0(M_{w,k} x_k + N_{w,k} T_{out} + p_{w,k} + N_{w,k}\xi_T) + T_{w,k,t=0,set}]/\alpha}] + \alpha\chi\} \leq 0 \quad (27)$$

where $P^Z_{sum} = \begin{bmatrix} P^Z_{sum,1} & P^Z_{sum,2} & \cdots & P^Z_{sum,H} \end{bmatrix}^T$ and has the same definition format for $\xi_P, \xi_T, \eta, \chi$. If the expectation expressions exist, then the minimization problem over $\alpha$ can be solved and these constraints are transformed to deterministic ones. Thus, the nominal distributions $P_{0t}, Q_{0t}$ need to take an exponential form.

Two ways are proposed here to obtain the nominal distribution according to the different methods of using the historical data. One is the Gaussian assumption (GA), in which it is assumed that the nominal distribution obeys a Gaussian distribution, the mean and variance of which are calculated from the first and second moments of the historical data, respectively. The other way is the KDE approximation (KDEA) [23], in which the uncertain variable $\xi$ has the following distribution function:
$$f_N(\xi) = \frac{1}{Nh_N}\sum_{i=1}^{N} H\left(\frac{\xi - \xi^i}{h_N}\right) \quad (28)$$

Here, $N$ is the amount of valid historical predicted-actual errors data $\xi^i$, $h_N$ is a positive constant number, and $H(\cdot)$ is a smooth function satisfying $H(\cdot) \geq 0$, $\int H(\xi)d\xi = 1$, $\int \xi H(\xi)d\xi = 0$, and $\int \xi^2 H(\xi)d\xi > 0$ [15]. We choose $H(\cdot)$ as a standard normal distribution, $N(0,1)$. Then, the $f_N(\xi)$ takes the sum of $N$ normal distributions, which have the same variance $h_N^2$ but a different mean $\xi^i$:
$$f_N(\xi) = \frac{1}{N}\sum_{i=1}^{N} \frac{1}{h_N\sqrt{2\pi}} e^{-\frac{(\xi-\xi^i)^2}{2h_N^2}} \quad (29)$$

It was shown in [24] that the KDE function converges to the real distribution in a 1-norm sense,
$$\int_{\mathbb{R}^K} |f_N(\xi) - f(\xi)|d\xi \to 0 \text{ as } N \to \infty \quad (30)$$

Thus, we reach the following formulations for constraints (24)–(27) and show an explanation for (24) here.

*(1) Gaussian Assumption (GA)*
Suppose $\xi_{P,t} \sim N(\mu_{P,t}, \sigma^2_{P,t})$, and for $\forall t = 1,2,\cdots,H$, (24) equals
$$\min_{\alpha\geq 0}\{\alpha \ln \int e^{[\sum_{k\in Z} x_{k,t} P_{HP,k} + P^Z_{sum,t} - P^Z_{max}]/\alpha} e^{\xi_{P,t}/\alpha} \frac{1}{\sigma_{P,t}\sqrt{2\pi}} e^{-\frac{(\xi_{P,t}-\mu_{P,t})^2}{2\sigma^2_{P,t}}} d\xi_{P,t}] + \alpha\eta_t\} \leq 0$$

Determining the minimum of $\alpha$, we have
$$\sum_{k\in Z} x_{k,t} P_{HP,k} + P^Z_{sum,t} - P^Z_{max} + \mu_{P,t} + \sigma_{P,t}\sqrt{2\eta_t} \leq 0 \quad (31),$$

which is a tightened deterministic constraint of (1). The derivation procedure is listed in Appendix B1.

*(2) KDE Approximation (KDEA)*
Under KDEA, for $\forall t = 1,2,\cdots,H$, the constraint (24) is transformed into
$$\sum_{k\in Z} x_{k,t} P_{HP,k} + P^Z_{sum,t} - P^Z_{max} + \min_{\alpha\geq 0}\{\alpha\eta_t + \frac{h^2_{N,P}}{2\alpha} + \alpha \ln \frac{1}{N_{P,t}}\sum_{i=1}^{N_{P,t}} e^{\frac{\xi^i_{P,t}}{\alpha}}\} \leq 0 \quad (32)$$

where $N_{P,t}$ is the amount of power forecast errors for time $t$ (details are shown in Appendix B2). We show that with given specific $\eta_t$ and $h_{N,P}$ and historical power forecast error data, $\xi^i_{P,t}$, the function $g_{P,t}(\alpha)$

$$g_{P,t}(\alpha) = \alpha\eta_t + \frac{h_{N,P}^2}{2\alpha} + \alpha \ln \frac{1}{N}\sum_{i=1}^{N} e^{\frac{\xi_{P,t}^i}{\alpha}}, \alpha > 0 \quad (33)$$

is a convex function (Appendix C). Then, we denote the minimum value of $g_{P,t}(\alpha)$ as $g_{P,t,\min}(\alpha)$ and take $g_{P,\min}(\alpha) = \begin{bmatrix} g_{P,1,\min}(\alpha) & g_{P,2,\min}(\alpha) & \cdots & g_{P,H,\min}(\alpha) \end{bmatrix}^T$. Thus, constraint (24) is equivalent to

$$\sum_{k \in Z} x_k P_{HP,k} - y_{sum} + p_{\max} + g_{P,\min}(\alpha) \le 0 \quad (34)$$

which is also a deterministic constraint.

Similarly, (25)–(27) also can be transformed to deterministic forms:

$$J_k x_k + K_k T_{out} + K_k g_{T,\min}(\alpha) + l_k - \overline{T_k} \le 0 \quad (35)$$

$$-J_k x_k - K_k T_{out} - K_k g_{T,\min}(\alpha) - l_k + \underline{T_k} \le 0 \quad (36)$$

$$Z_0(M_{w,k} x_k + N_{w,k} T_{out} + N_{w,k} g_{T,\min}(\alpha) + p_{w,k}) + T_{w,k,t=0,set} \le 0 \quad (37)$$

where $g_{T,\min}(\alpha) = \begin{bmatrix} g_{T,1,\min}(\alpha) & g_{T,2,\min}(\alpha) & \cdots & g_{T,H,\min}(\alpha) \end{bmatrix}^T$,

$$g_{T,t,\min}(\alpha) = \alpha\chi_t + \frac{h_{N,T}^2}{2\alpha} + \alpha \ln \frac{1}{N_{T,t}}\sum_{i=1}^{N_{T,t}} e^{\frac{\xi_{T,t}^i}{\alpha}}, \alpha > 0 \quad (38)$$

where $N_{T,t}$ represents the temperature forecast error.

The KL-DRO with KDEA methods can be reformulated as Model 3, shown below.

*Model 3: KL-DRO-KDEA*

min (6)
s.t. (9), (11)(12)(14), (34)-(37)
$x_{k,t} \in \{0,1\}$, $\forall t = 1, \cdots, H, \forall k = 1, 2, \cdots, N, \forall Z$

### C. Historical data utilization with information loss

It is clear that the knowledge of uncertainties, concentrated from the available data, dominates the quality of the DRO program. Based on information theory, the uncertain variables and the historical data could be regarded as an information source and many independent observations. We apply the minimum information loss theorem as

$$\min\{I_{loss} \mid f(\xi^b) \to P(\xi)\} = \min \frac{1}{B}\sum_{b=1}^{B} \ln \frac{P(\xi_b)}{f(\xi_b)} \quad (39)$$

where $P(\xi_b)$ and $f(\xi_b)$ are the integral probability and the dropped-into frequency of bin $b$, respectively, which means using frequency to estimate probability by depicting a histogram of all the $\xi_b$. The measured space is divided equally into $B$ bars. When $B \to \infty$, we have:

$$\min_{B \to \infty} \frac{1}{B}\sum_{b=1}^{B} \ln \frac{P(\xi_b)}{f(\xi_b)} = \int P(\xi) \ln \frac{P(\xi)}{f(\xi)} d\xi = D_{KL}(P(\xi), f(\xi)) \quad (40),$$

which indicates that if we use the histogram method to estimate the real distribution of the uncertain vectors, we would, at the same time, show the least information loss using the KL divergence. The KDE function is an approximation of the histogram with an appropriate $h_N$ and converges to the histogram when $N, B \to \infty$ [23], further demonstrating the superiority of this method.

### D. Risk meaning of the radius of an ambiguity set

The radius $\eta$ of the ambiguity set dominates the possible deviated degree of the nominal distribution and is obviously linked to the conservatism of the DRO program. In [18], it was shown that a EC-DRO is equivalent to a chance-constrained program, where radius $\eta$ reflects a decision maker's risk level, $\beta$, with equation $\beta = e^{-\eta}$. For example, when risk level $\beta = 0.1$, the radius, $\eta = 2.3026$. This provides a reference for selecting a suitable radius of our DRO model.

## IV. NUMERICAL TESTS

In this section, we compared performance between the conventional RO, the KL-DRO with Gaussian assumption (GA-DRO), and the KL-DRO with KDEA assumption (KDEA-DRO) via Monte Carlo simulations.

### A. Test settings

The test system includes 10 heterogeneous houses with HPs and water tanks; their parameters are listed in Table 1. The COP of the HPs and the water-to-house efficiency are set as constants: $\gamma_{COP} = 3$ and $\gamma_{w2h} = 1$. The comfort range for indoor temperature is set to be $[18, 24]°C$, and $S_0 = 60\ kW$. A finite difference model of the house [25] is used for the thermal simulations with 5 s for each step. The period number is 288 per day with 5 min for each time interval.

**Table 1. Parameters of 10 houses, tanks, and heat pumps**

| House | 1 | 2 | 3 | 4 | 5 | 6 | 7 | 8 | 9 | 10 |
|---|---|---|---|---|---|---|---|---|---|---|
| R (°C/kW) | 2.8 | 2.9 | 3.0 | 2.9 | 3.1 | 3.1 | 2.8 | 3.0 | 2.6 | 3.2 |
| C (kJ/°C) | 5.4 | 5.2 | 4.6 | 4.1 | 5.9 | 4.7 | 5.1 | 5.3 | 5.1 | 4.5 |
| $T_0$ (°C) | 19 | 20 | 21 | 20 | 19 | 19 | 20 | 21 | 20 | 19 |
| $R_w$ (°C/kW) | 2.2 | 2.4 | 2.5 | 2.6 | 2.4 | 2.8 | 2.4 | 2.6 | 2.6 | 2.5 |
| $C_w$ (kJ/°C) | 4.9 | 4.9 | 4.8 | 5.0 | 5.0 | 5.4 | 5.8 | 5.2 | 4.9 | 5.3 |
| $T_{w0}$ (°C) | 42 | 45 | 47 | 45 | 42 | 42 | 45 | 47 | 45 | 42 |
| $P_{HP}$ (kW) | 5 | 4.7 | 4.3 | 4.7 | 4.8 | 5 | 4.7 | 4.3 | 4.7 | 4.8 |

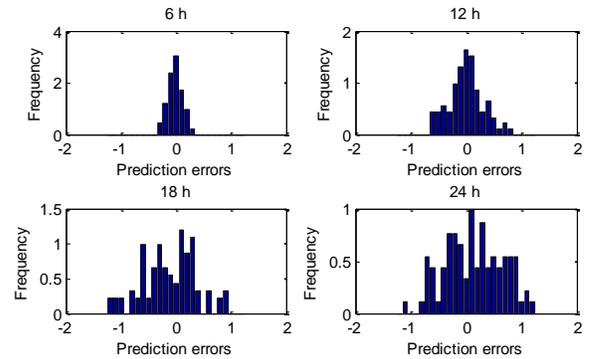

Figure 2. Histogram of 92 days' outdoor temperature prediction errors for four typical hours from day-ahead 23:30 to the coming day.

As shown in Figure 2, the day-ahead predicted errors of outdoor temperature were recorded for 92 days from November 1, 2016, to January 31, 2017, in Beijing, based on data in [26]. It can be seen that the early time nodes show relatively less forecast deviation. The prediction errors of the zonal power, the summation of PV, and load are generated randomly with an asymmetric $\chi^2$ distribution, $\xi_P = -0.15 rand(\chi_5^2) + 0.75$ for 5000 sample data for every moment. We tested the performance of different $h_{N,P}$ of the KDE function with power predicted errors as shown in

Figure 3. This shows that a larger $h_{N,P}$ may lead to a closer KDE function with more burr. We finally chose $h_{N,P}=0.2$ and $h_{N,T}=0.1$ for an appropriately smooth KDE function.

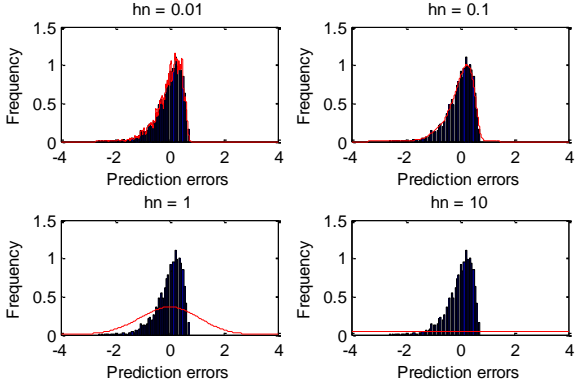

Figure 3. Histogram and KDE function of 5000 zonal power prediction errors

The predictions of PV power and the outdoor temperature curve are from real measurements on February 1, 2017, Tsinghua University, Beijing (Fig. 4). The TOU electricity price $\rho$ is shown in Figure 5; the mean value is $1\ \$/kWh$. The penalty weight for transformer capacity $\psi$ is $10\ \$/kW$ for the test cases.

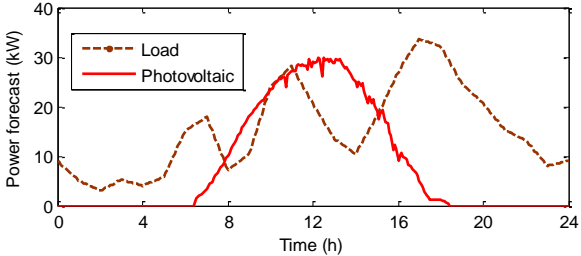

Figure 4. Day-ahead PV and load prediction

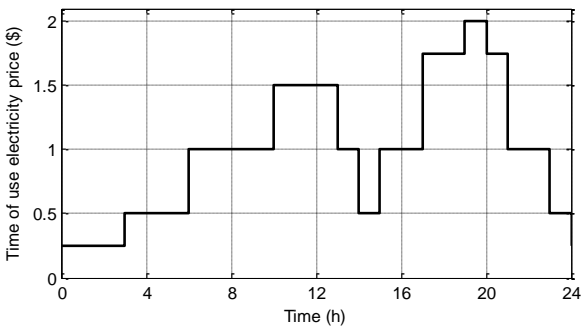

Figure 5. Curve of TOU electricity price

For the radius distance of the ambiguity set, we let $\eta_t = \chi_t = 2.3026, \forall t$ for a 0.9 confidence level. Numerical experiments were conducted in MATLAB 2014b with a core-i7 laptop. The minimization of $g(\alpha)$ was calculated by an inner-point method and the run time increased slightly when the historical data set increased. The mixed integer program was solved using CPLEX 12.5 with a 1% tolerance, and the average solving time for a KDEA-DRO problem was about 10 s.

### B. Comparison of KDEA-DRO with unscheduled conditions

For unscheduled HP operational conditions, the ON/OFF state of the HP obeys a simple hysteresis rule:

$$x_{k,t+1} = \begin{cases} 0, & if\ x_{k,t}=1\ \&\ T_{w,k,t} \geq \overline{T}_{w,k} \\ 1, & if\ x_{k,t}=0\ \&\ T_{w,k,t} \leq \underline{T}_{w,k} \\ x_{k,t}, & other\ conditions \end{cases} \quad (41)$$

The water temperature bounds $[\underline{T}_{w,k},\overline{T}_{w,k}]$ were set at $[40,45]\ °C$ here. The results for KDEA-DRO and unscheduled operation are shown in Figure 6 and Table 2. For the unscheduled scenario, Figure 6 shows that the transformer is overloaded from 17:30 to 19:30, when the electricity price is high. In contrast, the load profile of the transformer is flat and kept within its security limits with KDEA-DRO. Beyond that, HPs in the scheduled model turn off at high price moments to save money, and turn on at night for thermal energy storage, indicating good performance in load shifting. The large numerical differences of these models are shown in Table 3. The scheduled model cuts around 34% of the maximum daily transformer power and reduces the electricity cost for residents by about 18%.

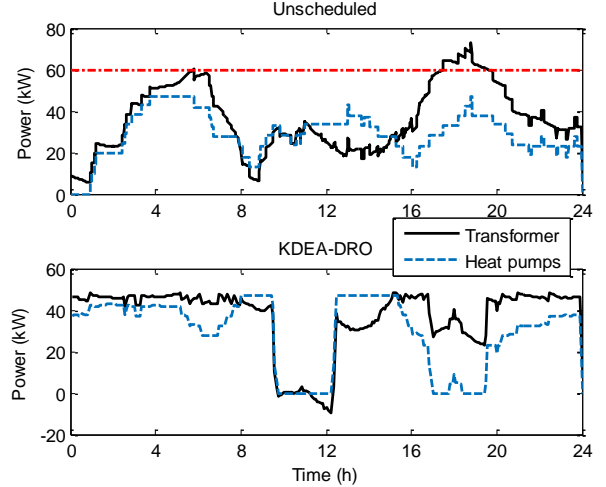

Figure 6. Operational power of transformer and heat pumps of the unscheduled model and KDEA-DRO model with $\psi = 10\ kW/\$$.

**Table 2. Main results of unscheduled operational model and KDEA-DRO model**

|  | $P_{max}$ (kW) | Peak cost ($) | Elec. cost ($) | Total ($) |
|---|---|---|---|---|
| Unscheduled | 73.204 | 732.0 | 962.7 | 1694.7 |
| KDEA-DRO | 48.612 | 486.1 | 787.6 | 1273.7 |

### C. Comparison of deterministic, KDEA-DRO, GA-DRO, and RO models

Operational strategies were generated using the deterministic, KDEA-DRO, GA-DRO, and RO models with the same setting and confirmed with 1000 Monte Carlo experiments. The intervals for RO were chosen with a 95% falling rate of





historical data and are symmetrical about zero (the concrete formulation of RO is shown in Appendix D). The forecast errors of outdoor temperature were generated by normally distributed random numbers, the means and variances of which were calculated from historical data, and the errors in power prediction were generated by the manipulated $\chi^2$ distribution mentioned above. The comfort rate (CR) is defined as the time between the comfort bounds divided by the total time.

$$CR_{day} = \min_{k}\left(\frac{Num(\underline{T_k} \leq T_{k,t} \leq \overline{T_k})}{Num(t)}\right) \qquad (42)$$

**Table 3. Results of Monte Carlo experiments for the operational strategies generated with different models**

| Method | | Deterministic | KDEA-DRO | GA-DRO | RO |
|---|---|---|---|---|---|
| $P_{max}$ /kW | Best day | 44.38 | 45.82 | 48.09 | 48.76 |
| | Worst day | 47.9 | 49.34 | 51.61 | 51.27 |
| | **Mean** | **47.18** | **48.62** | **50.89** | **50.56** |
| Elec cost /$ | Best day | 734.4 | 702.8 | 720.5 | 725.7 |
| | Worst day | 818.8 | 787.2 | 805 | 810.1 |
| | **Mean** | **801.7** | **770.1** | **787.8** | **793.0** |
| Comfort rate | Best day | 100.0% | 100.0% | 100.0% | 100.0% |
| | Worst day | 18.10% | 84.80% | 86.90% | 83.20% |
| | **Mean** | **85.6%** | **93.7%** | **95.8%** | **94.1%** |

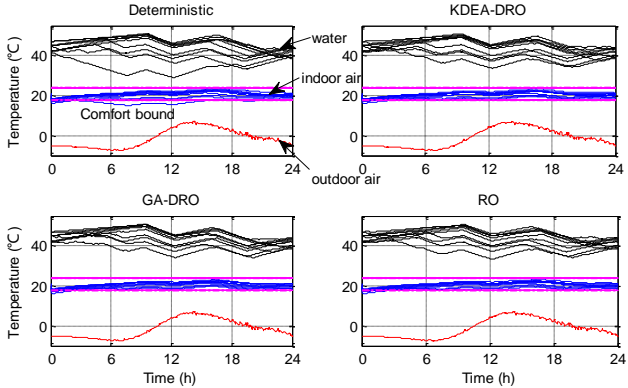

Figure 7. Temperature simulations of deterministic, KDEA-DRO, GA-DRO, and RO model for 10 houses

From Table 3, we can see that the DRO methods have relatively high comfort satisfaction with less energy cost than the RO method. Because the uncertainties are not considered, the deterministic method has a poorer comfort rate and a higher cost. Among the two DRO methods, $P_{max}$ and the energy cost with KDEA-DRO are less than with GA-DRO, because the KDE function is more accurate than the Gaussian function in representing historical information, while the comfort satisfaction of KDEA-DRO is acceptable. Regarding the indoor comfort aspect, outage of indoor temperature of one house can be observed in the deterministic curves in Figure 7, while the robustness is better in the three robust methods.

*D. Discussion: Influence of the divergences $\eta_t$ and $\chi_t$*

The ambiguity sets of the probability distribution of the uncertainties in our KDEA-DRO model are controlled by the divergences $\eta_t$ and $\chi_t$. As discussed in Section III, for an ambiguity set, the risk level $\beta$ and the set's divergence $\eta$, it holds that $\beta = e^{-\eta}$. When the risk level increases to one, the divergence decreases to zero, and the DRO problem reduces to a stochastic optimization problem.

We first analyze the $g_{P,\min}(\alpha)$ and four typical hours' $g_{T,t,\min}(\alpha)$, which represent a robust part in the KDEA-DRO model with different divergence settings. As shown in Figure 8, with the increase in divergence and the decrease in risk level, the $g_{P,\min}(\alpha)$ and $g_{T,t,\min}(\alpha)$ increase with a gradually decreasing rate due to the robustness of the relative constraints. Due to the higher forecast accuracy, the $g_{T,6h,\min}(\alpha)$ is smaller than at the three other times (Fig. 8).

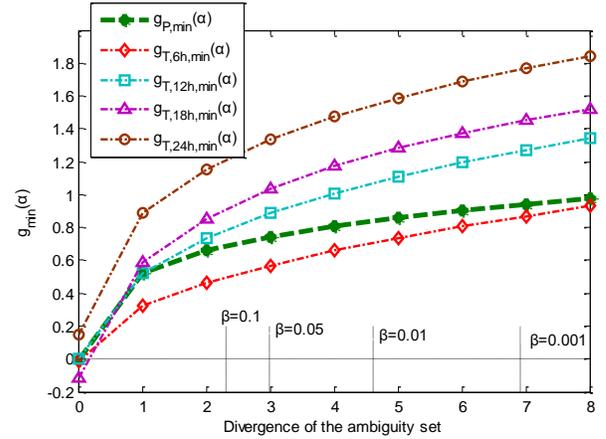

Figure 8. $g_{\min}(\alpha)$ of power and four typical hours prediction error data with different divergence η of the ambiguity set

Next, we discuss the influence of divergences $\eta_t$ and $\chi_t$ on the results of KDEA-DRO via 1000 time Monte Carlo simulations; the other parameters are the same as those in subsection *C*. We select four risk levels. β = 0.1, 0.05, 0.01, and 0.001, for the power and temperature ambiguity sets for comparison; the outcomes are shown in Table 4. With the increase in power divergence, $\eta_t$, the power peak increases, the electricity cost decreases, and the comfort rate rises. Thus, a large power divergence could provide an advantage in reducing the electricity cost and improving the users' thermal satisfaction. Moreover, with an increase in outdoor temperature divergence, $\chi_t$, the power peak increases slightly, the electricity cost increases, and the comfort rate rises. Based on these results, we can control the robustness and conservatism of the KDEA-DRO by varying the divergence settings.

**Table 4. Results of the KDEA-DRO under different divergences of $\eta_t$ and $\chi_t$ for different risk levels β**

(a) $P_{max}$ (kW)

| $\chi_t$ (risk level) \ $\eta_t$ (risk level) | 2.3026 (β = 0.1) | 2.9957 (β = 0.05) | 4.6052 (β = 0.01) | 6.9078 (β = 0.001) |
|---|---|---|---|---|
| 2.3026 (β = 0.1) | 48.62 | 49.51 | 52.11 | 52.21 |
| 2.9957 | 48.37 | 49.42 | 51.27 | 52.37 |

| | | | | |
|---|---|---|---|---|
| (β = 0.05) | | | | |
| 4.6052 (β = 0.01) | 49.51 | 50.65 | 51.46 | 52.56 |
| 6.9078 (β = 0.001) | 50.63 | 51.41 | 51.76 | <u>52.89</u> |

**(b) Electricity cost ($)**

| $\chi_t$ (risk level) \ $\eta_t$ (risk level) | 2.3026 (β = 0.1) | 2.9957 (β = 0.05) | 4.6052 (β = 0.01) | 6.9078 (β = 0.001) |
|---|---|---|---|---|
| 2.3026 (β = 0.1) | 770.1 | 768.6 | 765.4 | <u>755.0</u> |
| 2.9957 (β = 0.05) | 777.0 | 784.9 | 771.1 | 758.0 |
| 4.6052 (β = 0.01) | 784.5 | 785.2 | 773.7 | 763.8 |
| 6.9078 (β = 0.001) | <u>788.5</u> | 783.5 | 776.6 | 764.8 |

**(c) Comfort rate**

| $\chi_t$ (risk level) \ $\eta_t$ (risk level) | 2.3026 (β = 0.1) | 2.9957 (β = 0.05) | 4.6052 (β = 0.01) | 6.9078 (β = 0.001) |
|---|---|---|---|---|
| 2.3026 (β = 0.1) | <u>93.70%</u> | 95.47% | 95.78% | 96.02% |
| 2.9957 (β = 0.05) | 94.90% | 95.52% | 96.28% | 96.31% |
| 4.6052 (β = 0.01) | 95.13% | 95.68% | 96.12% | 96.42% |
| 6.9078 (β = 0.001) | 95.71% | 95.71% | 96.18% | <u>96.50%</u> |

## V. CONCLUSIONS AND FUTURE WORK

In this paper, we developed a KL distance-based DRO model of HP-DOSM with residential temperature constraints, which can both decrease the peak-valley gap and the cost to residents under a TOU electricity price. This distance-based DRO model can well capture the uncertainties of weather prediction, photovoltaic prediction, and load prediction errors, while it is tractable. Numerical tests showed that our distance-based DRO was robust with less conservatism than the conventional RO model. Moreover, the robustness of this model can be adjusted by tuning the risk level, which has an explicit meaning in the optimization problem. In future work, a distributed algorithm is needed to solve HP-DOSM for large-scale distribution networks integrated with massive heat pumps.

## APPENDIX

### A. Detailed structural transformation of the heat pump day-ahead operational schedule model

We first apply several simplifications. Transform the (3) and (4) equalities to a simplified form:

$$T_{k,t+1} = \delta_k T_{k,t} + \varphi_k T_{w,k,t} + \gamma_k T_{out,t} \tag{43}$$

$$T_{w,k,t+1} = \delta_{w,k} T_{k,t} + \varphi_{w,k} T_{w,k,t} + \lambda_{w,k} x_{k,t} \tag{44}$$

With the definition of the vectors in the nomenclature, we have:

$$\begin{bmatrix} 1 \\ -\delta_k & 1 \\ & \ddots & \ddots \\ & & -\delta_k & 1 \end{bmatrix} T_k = \begin{bmatrix} \varphi_k \\ & \ddots \\ & & \varphi_k \end{bmatrix} T_{w,k} + \begin{bmatrix} \gamma_k \\ & \ddots \\ & & \gamma_k \end{bmatrix} T_{out} + \begin{bmatrix} T_{k,t=0,set} \\ \\ \\ \end{bmatrix} \tag{45}$$

This can be rewritten as

$$A_k T_k = B_k T_{w,k} + C_k T_{out} + d_k \tag{46}$$

In the same way, we can state that

$$E_{w,k} T_{w,k} = F_{w,k} T_k + G_{w,k} x_k + h_{w,k} \tag{47}$$

It is easy to show that $A_k$ and $E_{w,k}$ are non-singular; thus, we have

$$(A_k - B_k E_{w,k}^{-1} F_{w,k}) T_k = B_k E_{w,k}^{-1} G_{w,k} x_k + C_k T_{out} + B_k E_{w,k}^{-1} h_{w,k} + d_k \tag{48}$$

$$(E_{w,k} - F_{w,k} A_k^{-1} B_k) T_{w,k} = F_{w,k} A_k^{-1} C_k T_{out} + G_{w,k} x_k + h_{w,k} + F_{w,k} A_k^{-1} d_k \tag{49}$$

Due to the Schur complement, the matrices $(A_k - B_k E_{w,k}^{-1} F_{w,k})$ and $(E_{w,k} - F_{w,k} A_k^{-1} B_k)$ are non-singular, so we have:

$$T_k = (A_k - B_k E_{w,k}^{-1} F_{w,k})^{-1} (B_k E_{w,k}^{-1} G_{w,k} x_k + C_k T_{out} + B_k E_{w,k}^{-1} h_{w,k} + d_k) \tag{50}$$

$$T_{w,k} = (E_{w,k} - F_{w,k} A_k^{-1} B_k)^{-1} (F_{w,k} A_k^{-1} C_k T_{out} + G_{w,k} x_k + h_{w,k} + F_{w,k} A_k^{-1} d_k) \tag{51}$$

represented by

$$T_k = J_k x_k + K_k T_{out} + l_k \tag{52}$$

$$T_{w,k} = M_{w,k} x_k + N_{w,k} T_{out} + p_{w,k} \tag{53}$$

where $T_k$ and $T_{w,k}$ are decoupled.

### B. Detailed reformulation of Gaussian Assumption and KDE Approximation in constraint (24)

*B.1 Gaussian Assumption*

The inner formula of (24) equals

$$\alpha \ln \int e^{[\sum_{k \in Z} x_{k,t} P_{HP,k} + P_{sum,t}^Z - P_{max}^Z + \xi_{P,t}]/\alpha} \frac{1}{\sigma_{P,t} \sqrt{2\pi}} e^{-\frac{(\xi_{P,t} - \mu_{P,t})^2}{2\sigma_{P,t}^2}} d\xi_{P,t}] + \alpha \eta_t$$

$$= \sum_{k \in Z} x_{k,t} P_{HP,k} + P_{sum,t}^Z - P_{max}^Z + \mu_{P,t} + \alpha \eta_t + \frac{\sigma_{P,t}^2}{2\alpha} \tag{54}$$

Due to $\min_{\alpha \geq 0} \{\alpha \eta_t + \frac{\sigma_{P,t}^2}{2\alpha}\} = \sigma_{P,t} \sqrt{2\eta_t}$, we take (24) as

$$\sum_{k \in Z} x_{k,t} P_{HP,k} + P_{sum,t}^Z - P_{max}^Z + \mu_{P,t} + \sigma_{P,t} \sqrt{2\eta_t} \leq 0 \tag{55}$$

*B.2 KDE Approximation*

The inner formula of (24) equals

$$\alpha \ln \int e^{[\sum_{k \in Z} x_{k,t} P_{HP,k} + P_{sum,t}^Z - P_{max}^Z + \xi_{P,t}]/\alpha} \frac{1}{N_{P,t}} \sum_{i=1}^{N_{P,t}} \frac{1}{h_{N,P} \sqrt{2\pi}} e^{-\frac{(\xi_{P,t} - \xi_{P,t}^i)^2}{2h_{N,P}^2}} d\xi_{P,t}] + \alpha \eta_t$$

$$= \sum_{k \in Z} x_{k,t} P_{HP,k} + P_{sum,t}^Z - P_{max}^Z + \alpha \eta_t + \frac{h_{N,P}^2}{2\alpha} + \alpha \ln \frac{1}{N_{P,t}} \sum_{i=1}^{N_{P,t}} e^{\frac{\xi_{P,t}^i}{\alpha}} \tag{56}$$

### C. Convexity proof of $g(\alpha)$

The convexity of $g(\alpha)$ can be calculated by a second order differential function:

$$g(\alpha) = \alpha \eta + \frac{h_N^2}{2\alpha} + \alpha \ln \frac{1}{N} \sum_{i=1}^{N} e^{\frac{\xi^i}{\alpha}}, \alpha > 0 \tag{57}$$



$$g'(\alpha) = \eta - \frac{h_N^2}{2\alpha^2} + \ln\left[\frac{\sum_{i=1}^{N} e^{\frac{\xi^i}{\alpha}}}{N}\right] + \frac{\alpha \sum_{i=1}^{N} -\frac{e^{\frac{\xi^i}{\alpha}}\xi^i}{\alpha^2}}{\sum_{i=1}^{N} e^{\frac{\xi^i}{\alpha}}} \quad (58)$$

$$g''(\alpha) = \frac{h_N^2}{\alpha^3} + \frac{\left(\sum_{i=1}^{N} e^{\frac{\xi^i}{\alpha}} \xi^i\right)^2}{\left(\sum_{i=1}^{N} e^{\frac{\xi^i}{\alpha}}\right)^2} + \frac{\sum_{i=1}^{N} \frac{e^{\frac{\xi^i}{\alpha}}(\xi^i)^2}{\alpha^3}}{\sum_{i=1}^{N} e^{\frac{\xi^i}{\alpha}}} > 0 \quad (59)$$

Thus, $g(\alpha)$ is convex.

*D. Robust optimization model*

$$\min_{s.t.} \, \max_{\xi_P \in [\underline{\xi}_P, \overline{\xi}_P]} \sum_{k \in Z} x_k P_{HP,k} + P_{sum}^Z - P_{max}^Z + \xi_P \leq 0 \quad (60)$$

$$\max_{\xi_T \in [\underline{\xi}_T, \overline{\xi}_T]} J_k x_k + K_k T_{out} + l_k - \overline{T}_k + K_k \xi_T \leq 0 \quad (61)$$

$$\max_{\xi_T \in [\underline{\xi}_T, \overline{\xi}_T]} -J_k x_k - K_k T_{out} - l_k + \underline{T}_k - K_k \xi_T \leq 0 \quad (62)$$

$$\max_{\xi_T \in [\underline{\xi}_T, \overline{\xi}_T]} Z_0 (M_{w,k} x_k + N_{w,k} T_{out} + p_{w,k} + N_{w,k} \xi_T) + T_{w,k,t=0,set} \leq 0 \quad (63)$$

$(9), (11)(12)(14)$
$x_{k,t} \in \{0,1\}$, $\forall t = 1, \cdots, H, \forall k = 1, 2, \cdots, N, \forall Z$